\documentclass{ws-ijmpd}
\usepackage[super,compress]{cite}
\usepackage{graphicx}
\begin{document}

\markboth{M. H. Chan}
{Analytic expressions for the dark matter-baryon relations}
%
\catchline{}{}{}{}{}
%

\title{Analytic expressions for the dark matter-baryon relations}
\author{Man Ho Chan}
\address{Department of Science and Environmental Studies,
\\ The Education University of Hong Kong,
\\  Tai Po, New Territories, Hong Kong, China
\\ chanmh@eduhk.hk}
 
\maketitle

\begin{history}
\received{Day Month Year}
\revised{Day Month Year}
\end{history}

\begin{abstract}
Recently, some very strong correlations between the distribution of dark matter and baryons (the dark matter-baryon relations) in galaxies with very different morphologies, masses, sizes, and gas fractions have been obtained. Some models have been suggested to explain why the dark matter contribution is fully specified by that of the baryons. In this article, we derive two analytic expressions to explain the observed dark matter-baryon relations based on the cold dark matter (CDM) model. The resultant expressions give excellent agreement with the observational data. The parameters involved in the analytic expressions are closely related to the amount of the baryon content. This model can provide a theoretical understanding of the strong correlations observed. We suggest that the observed relation represents the end product of galaxy formation.
\end{abstract}


\section{Introduction}
Recent empirical fits indicate a very strong correlation (with very small scatter) between the radial acceleration traced by rotation curves ($g_t=v^2/r=|\partial \Phi/\partial r|$, where $\Phi$ is the total gravitational potential and $v$ is the rotational velocity) and the radial acceleration predicted by the observed distribution of baryons ($g_b=|\partial \Phi_b/\partial r|$, where $\Phi_b$ is the gravitational potential of the baryonic component) for 153 rotationally supported galaxies \cite{McGaugh}. These galaxies have different morphologies, masses, sizes, and gas fractions. Prior to this finding, another very strong correlation between the central surface density of stars $\Sigma_*^0$ and dynamical mass $\Sigma_D^0$ in 135 disk galaxies has been obtained \cite{Lelli}. These results indicate that there exists a strong connection between baryon and dark matter distribution. They are also closely related to some other relations between dark matter and baryons such as the baryonic Tully-Fisher relation \cite{Tully,McGaugh2,Lelli2} and the `Halo-Disk' conspiracy problem \cite{Battaner,Remus}. It seems that the dark matter contribution is fully specified by that of the baryons. 

The radial acceleration relation can be well described by the following function \cite{McGaugh}
\begin{equation}
g_t=\frac{g_b}{1-e^{-\sqrt{g_b/g_0}}},
\end{equation}
where $g_0=1.20\pm 0.02(\rm random)\pm 0.24 (\rm systematic) \times 10^{-10}$ m/s$^2$, and the central-surface-densities relation can be described by a double power law \cite{Lelli}
\begin{equation}
\Sigma_D^0=\Sigma_0 \left[1+ \frac{\Sigma_*^0}{\Sigma_{\rm crit}} \right]^{\alpha-\beta} \left[ \frac{\Sigma_*^0}{\Sigma_{\rm crit}} \right]^{\beta},
\end{equation}
where $\alpha$, $\beta$, $\Sigma_{\rm crit}$ and $\Sigma_0$ are fitted parameters. In fact, these two relations are closely related because one can relate the central surface density with the radial acceleration by $\Sigma_D^0=(2\pi G)^{-1}g_t$. In general, both relations give a linear slope at high accelerations (high baryonic surface density) and $g_t \propto \sqrt{g_b}$ at low accelerations (low baryonic surface density).

Based on these findings, Milgrom (2016) \cite{Milgrom} shows that the Modified Newtonian Dynamics (MOND) can give a satisfactory explanation to the strong correlations. On the other hand, some studies try to use semi-empirical model to explain a similar relation - the mass discrepancy acceleration relation (MDAR) \cite{Cintio}. However, the resulting correlation involves some model-dependent parameters and universal forms of baryon distribution, which might not be strong enough to explain the new correlations. Recently, Desmond (2017) \cite{Desmond} uses a statistical way (the abundance-matching paradigm) and shows that cold dark matter (CDM) model can also account for the strong correlations. In this article, we use another approach and derive analytic expressions to explain these relations based on the CDM model. Our results can give excellent agreements with the observed data, within a very small error bars. We show that the parameters involved in the expressions are closely related to the amount of the baryon content. It can explain why the observed relations are so tight. Based on the sample used in McGaugh \& Lelli (2016) \cite{McGaugh}, the spiral galaxies can be roughly classified as three different types: bulge-dominated galaxies (BDG), disk-dominated galaxies (DDG) and gas-dominated galaxies (GDG). We will derive the corresponding analytic expression for each type of galaxies.

\section{The dark matter-baryon relation for bulge-dominated galaxies}
Assume that dark matter would form structure first in galaxy formation. The distribution of the baryonic component would be affected by the dark matter distribution via gravitational interaction. In fact, baryonic processes might affect dark matter distribution near the central part of some galaxies. Nevertheless, in general, baryonic matter has only a minor effect on dark matter distribution, especially in large $r$ region \cite{Martinsson,Pontzen}. In CDM scenario, dark matter particles interact each other through gravity only. Numerical simulations show that dark matter density follows the Navarro-Frenk-White (NFW) density profile \cite{Navarro}
\begin{equation}
\rho_d=\frac{\rho_sr_s}{r} \left(1+ \frac{r}{r_s} \right)^{-2},
\end{equation}
where $\rho_s$ and $r_s$ are the scale density and scale radius respectively. The integrated mass profile is $M_d=4\pi \rho_sr_s^3[\ln (1+r/r_s)-r/(r_s+r)]$. This universal profile gives good agreements in many galaxies and clusters \cite{Pointecouteau,Breddels,Viola,Sofue}, including our Milky Way \cite{Iocco,Pato}. Although some observations indicate that the dark matter density of the inner regions of many galaxies should be cored \cite{deBlok}, this would just contribute a small error as baryons usually dominate the inner regions of most galaxies. Therefore, using the NFW profile is still a very good choice in this analysis. By using the NFW profile, the radial acceleration due to dark matter gravity is:
\begin{equation}
g_d=4\pi \rho_sr_sG \left( \frac{r_s}{r} \right)^2 \left[\ln \left(1+\frac{r}{r_s} \right)-\frac{r}{r+r_s} \right].
\end{equation}

For BDG, the bulge contribution dominates the baryonic matter contribution of rotational velocity for a large range of $r$ \cite{McGaugh}. We can approximate this contribution by
\begin{equation}
g_b \approx \frac{GM_b}{r^2}=g_{b0} \left(\frac{r_s}{r} \right)^2,
\end{equation}
where $M_b$ is the total mass of the bulge and $g_{b0}=GM_b/r_s^2$. By writing the total radial acceleration $g_t=g_b+g_d$ and putting Eq.~(5) into Eq.~(4), we get
\begin{equation}
g_t=g_b+g_b \left( \frac{g_{d0}}{g_{b0}} \right) \left[ \ln \left(1+\sqrt{\frac{g_{b0}}{g_b}} \right)-\frac{\sqrt{g_{b0}/g_b}}{1+\sqrt{g_{b0}/g_b}} \right],
\end{equation}
where $g_{d0}=4\pi \rho_sr_sG$. The above simple relation is the analytic expression for BDG. For a typical BDG, the value of $g_b$ is ranging from $10^{-11}$ m/s$^2$ to $10^{-8}$ m/s$^2$ for different $r$. In Fig.~1, we use Eq.~(6) and plot $g_t$ against this range of $g_b$. A very good fit can be obtained when $g_{d0}/g_{b0}=4$ and $g_{b0}=10^{-10}$ m/s$^2$. However, for some other values of $g_{d0}/g_{b0}$ (e.g. $g_{d0}/g_{b0}=2$), the fit is quite poor (see also Fig.~1). It seems that there exists some fine tuning in the ratio $g_{d0}/g_{b0}$ and the value $g_{b0}$. 

Nevertheless, these values are determined by some other factors. For $g_{d0}=4\pi\rho_sr_sG$, this value depends on the total mass of dark matter because $\rho_sr_s \propto M_{dt}^{0.2}$, where $M_{dt}$ is the total dark matter mass in BDG. In the CDM model, we have $\rho_s=200\rho_cc^3/3f(c)$ and $r_s=(3M_{dt}/800\pi \rho_cc^3)^{1/3}$, where $c \sim 4-40$ is the concentration parameter, $\rho_c=9\times 10^{-30}$ g cm$^{-3}$ is the cosmological critical density and $f(c)=\ln(1+c)-c/(c+1)$ \cite{Navarro}. Therefore, we can get $\rho_sr_s \propto c^{1.36}M_{dt}^{1/3}$ as it can be shown that $c^2/f(c) \propto c^{1.36}$ for $c=4-40$. Based on the simulation results for the CDM model, we have $c=5.05(M_{dt}/10^{14}h^{-1}M_{\odot})^{-0.101}$, where $h \approx 0.7$ is the Hubble parameter \cite{Schaller}. Therefore, we get $\rho_sr_s \approx 144(M_{dt}/10^{12}M_{\odot})^{0.2}~M_{\odot}$ pc$^{-2}$ $\propto M_{dt}^{0.2}$. This result agrees with empirical observations \cite{DelPopolo}. For a BDG, the typical value of $M_{dt} \sim 10^{13}M_{\odot}$ \cite{Lelli3} gives $g_{d0} \approx 4 \times 10^{-10}$ m/s$^2$. Since $\rho_sr_s$ depends on $M_{dt}$ slowly, the actual range of $g_{d0}$ is very small. Furthermore, the ratio of the total baryonic mass to total dark matter mass can be written as 
\begin{equation}
\frac{M_b}{M_{dt}}=\frac{g_{b0}}{4\pi G \rho_sr_sf(c)}=\frac{g_{b0}}{g_{d0}f(c)}.
\end{equation}
For a BDG, we have $c=6-7$ and $f(c) \approx 1.2$. Therefore, we get $M_b/M_{dt}=0.2$, which is same as the value $0.2$ (baryon to dark matter ratio) predicted from standard cosmology. In other words, if $M_b/M_{dt}$ is close to 0.2 and the allowed ranges of the values $\rho_sr_s$ and $f(c)$ are small for all BDG, the possible range of $g_{b0}$ is also small. In fact, since BDG are large galaxies, the ratio $M_b/M_{dt}$ would be quite close to the cosmological value. In our empirical fits, we have $g_{b0}/g_{d0}=4$ and $g_{b0}=10^{-10}$ m/s$^2$. This would give $g_{d0}=4 \times 10^{-10}$ m/s$^2$, which is same as the CDM model's prediction. Therefore, the CDM theory can explain why the ratio $g_{b0}/g_{d0}$ and the value of $g_{b0}$ are somewhat `fine-tuned'. This also explains why the observed radial acceleration relation is so tight (the error bars are very small).

Besides the acceleration relation, we can also fit our expression with the central-surface-density relation \cite{Lelli} (see Fig.~2). The result is in good agreement with the observed data.

\section{The dark matter-baryon relation for disk-dominated galaxies and gas-dominated galaxies}
The gravitational effect of dark matter for baryonic matter in DDG and GDG can be analyzed by the steady-state Jeans equation \cite{Evans}:
\begin{equation}
\frac{d(\rho_b \sigma_b^2)}{dr}=- \rho_b \frac{\partial \Phi}{\partial r},
\end{equation}
where $\rho_b$ is the baryonic mass density and $\sigma_b$ is the radial velocity dispersion of baryonic matter. Although the above equation assumes spherically symmetric, we can still apply it in cylindrical disk-like case as we mainly focus on the data near $z=0$. Here, $r$ is the radius in cylindrical coordinate. For the DDG and GDG, we assume that the baryonic mass density follows a 2D-like disk and goes like $\rho_b(r) \propto r^{-\gamma}$, where $0< \gamma< 3$ is a constant. Therefore, the baryonic mass function can be simply given by $M_b(r) \approx C\rho_b(r)r^2z_0$, where $z_0$ is the scale height of the disk and $C$ is a constant which depends on the functional form of $\rho_b(r)$. Therefore, we have $\rho_b(r)=g_b/GCz_0$. Also, in a rotationally supported galaxy, the baryonic velocity dispersion is approximately given by $\sigma_b \approx C' \sqrt{g_tr}$, where $C' \approx 0.7-1$. Putting these relations and $g_t=\partial \Phi/\partial r$ into Eq.~(8), we have 
\begin{equation}
C'^2 \frac{d}{dr}(g_bg_tr)=-g_bg_t.
\end{equation}
By writing $u=g_bg_t$, we can get
\begin{equation}
\frac{d \ln u}{d \ln r}=-K,
\end{equation}
where $K=1+C'^{-2}$. Integrating the above equation, we can get
\begin{equation}
g_bg_t=C"\left(\frac{r}{r_s} \right)^{-K},
\end{equation}
where $C"$ is a constant which depends on the baryonic content of a galaxy. This is the fundamental equation to relate $g_b$ with $g_t$ for the DDG and GDG. Since $g_t=g_b+g_d$, we can finally get
\begin{equation}
g_b=\frac{-g_d+\sqrt{g_d^2+4C"(r/r_s)^{-K}}}{2}.
\end{equation}
The relation in Eq.~(12) can be obtained by using the NFW density profile for $g_d$ in Eq.~(4). The value of $\rho_sr_s$ and the constant $C"$ determine the functional form of the relation. Generally speaking, different values of $y=r/r_s$ would give different values of $g_b$ and $g_t$. For DDG and GDG, the typical ranges of $g_b$ and $M_{dt}$ are $10^{-12}-10^{-10}$ m/s$^2$ and $10^{11}-10^{12}M_{\odot}$ respectively \cite{Lelli3}. Using $\rho_sr_s=144(M_{dt}/10^{12}M_{\odot})^{0.2}M_{\odot}$ pc$^{-2}$, the value of $g_{d0}$ is about $2 \times 10^{-10}$ m/s$^2$. Taking $C'=0.7$ ($K=3$), we plot the radial acceleration relation in Fig.~1. We can see that we can obtain a very good fit with these parameters. Similar good fits can also be obtained if we use $C'=1$ ($K=2$). The resulting relation does not sensitively depend on $C'$ for DDG and GDG. Here, we define a new parameter $z \equiv g_b/g_{d0}$ at $y=5$ to represent the constant $C"$ and the total baryonic content of a galaxy. Good fits can be obtained for a wide range of $z=0.05-0.18$, which means about 85\%-95\% of mass is dark matter. This is consistent with observations. Since DDG and GDG are small structures that may be formed due to fragmentation of large structures, the possible range of baryonic content in DDG and GDG is much larger than that in BDG. Our results are consistent with this prediction. Nevertheless, since the radial acceleration relation does not sensitively depend on $z$, the wide range of baryonic content (represented by $z$) can give a tight dark matter-baryonic relation for $g_b=10^{-12}-10^{-10}$ m/s$^2$.

We also plot the central-surface-density relation for the DDG and GDG in Fig.~2. We can obtain excellent agreements with both dark matter-baryon relations for these galaxies. 

\begin{figure}
\vskip 10mm
 \includegraphics[width=82mm]{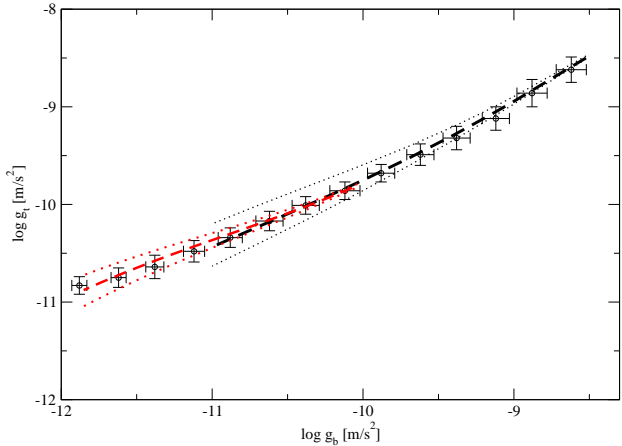}
 \caption{The resulting radial acceleration relation ($\log g_t$ vs. $\log g_b$) and the observed data with error bars \cite{McGaugh}. The black dashed line represents the best-fit relation for the BDG. The red dashed line represents the best-fit relation for the DDG and GDG. The black dotted lines represent the relations with $g_{d0}=2\times 10^{-8}$ m/s$^2$ and $g_{d0}=8\times 10^{-8}$ m/s$^2$. The red dotted lines represent the relations for $z=0.05$ and $z=0.18$.} 
\vskip 10mm
\end{figure}

\begin{figure}
\vskip 10mm
 \includegraphics[width=82mm]{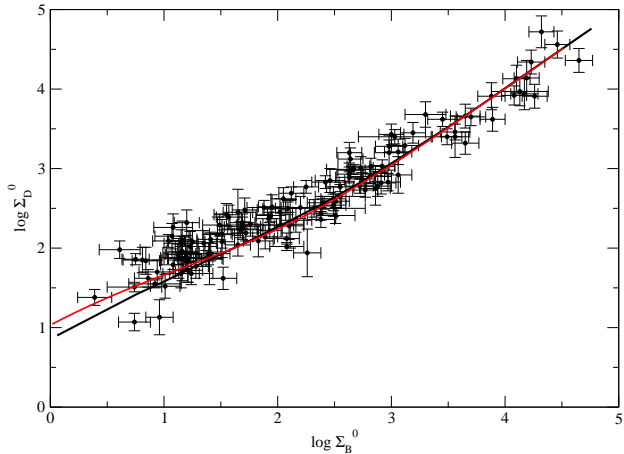}
 \caption{The resulting central-surface-densities relation ($\log \Sigma_D^0$ vs. $\log \Sigma_B^0$, where $\Sigma_B^0$ is the central surface density of baryons) and the observed data with error bars \cite{Lelli}. Here, we assume that $\Sigma_*^0$ is the proxy for $\Sigma_B^0$ ($\Sigma_*^0=\Sigma_B^0$) \cite{Milgrom} and the unit of the surface central densities are $M_{\odot}$ pc$^{-2}$. The black solid line: the best-fit relation for BDG. The red solid line: the best-fit relation for DDG and GDG.} 
\vskip 10mm
\end{figure}

\section{Discussion}
In this article, we derive the radial acceleration relation by using the CDM framework. Our results give excellent agreements with the observed data. Our model can also explain for the observed central-surface-densities relation in disk galaxies. The fitted parameters are in good agreement with the prediction in the CDM model, which support the findings in Desmond (2017) \cite{Desmond}. The analytic expressions derived in this article can provide a theoretical understanding of the results obtained in Desmond (2017) and give another supporting evidence for the CDM paradigm. Generally speaking, the functional form of the relations in BDG, DDG and GDG is controlled by two parameters: the value of $\rho_sr_s$ and the ratio of total baryonic mass to total dark matter mass. The CDM models suggest that $\rho_sr_s$ slowly depends on the total dark matter mass $M_{dt}$. Therefore, the range of the value $\rho_sr_s$ is small for all galaxies. For the ratio of total baryonic mass to total dark matter mass, different morphologies would have different ratios. For BDG, the ratio is about 0.2, which is close to the cosmological ratio. For DDG and GDG, the corresponding ratio is about 0.05-0.18. This suggests that DDG and GDG are rich in dark matter ($>85-95$\% is dark matter). In fact, observations indicate that the bulgeless galaxies (DDG) and dwarf galaxies (GDG) are dark matter dominated \cite{Bertone}. Therefore, our results give a consistent picture in the CDM model and observations. Furthermore, our result in Eq.~(12) suggests some universal forms of baryonic distribution. Interestingly, recent findings indicate that the rotation curves in specifically normalized units look all alike \cite{Karukes}, which is consistent with our result.

Overall speaking, there are three assumptions in our model. The first assumption is that the baryonic density distribution for DDG and GDG is determined by the steady-state Jeans equation, which can be derived from the general collisionless Boltzmann equation. It assumes that there is no interaction between dark matter particles and baryons, and the dark matter and baryonic density distribution is in equilibrium state. Some of the recent studies start to investigate the accuracy of using Jeans model to estimate the dynamical mass of low-mass galaxies \cite{Li,Badry}. Although simulations show that the starburst and stellar feedback might affect the dynamical mass estimation, these outflows are not large enough to introduce systematic errors in the estimation using the Jeans model \cite{Read}. Also, if galaxies completely lose their gas, the Jeans model would still be reliable to model galaxies \cite{Badry}. Therefore, the observed tight acceleration relation might show that most of the galaxies have already entered the final stage of galaxy formation. The second assumption is that the baryonic density profile in DDG and GDG follows a simple functional form, $\rho_b \propto r^{-\gamma}$. Although we usually model baryonic disks by exponential functions, this is also a good assumption because the functional behaviors between $\rho_b \propto r^{-\gamma}$ and an exponential function are similar when $r$ is large. However, the bulge contribution for DDG and the disk contribution for BDG are neglected in our model. Based on the sample used in McGaugh \& Lelli (2016) \cite{McGaugh}, a few DDG have small bulges in the inner part and some BDG have small disks in the outer region. Therefore, our result may have a small change if these effects are taken into account. The third assumption is that we use the NFW profile to model the dark matter density profile. Although some studies point out that the NFW profile is not a good profile to model some of the small galaxies \cite{deBlok}, especially for some dwarf galaxies \cite{Burkert}, it is the most commonly used profile to model the CDM particles \cite{Iocco}, such as modeling dark matter annihilation \cite{Ackermann}. It is supported by numerical simulations and observations in many galaxies and clusters. Latest simulation results also indicate that the CDM model works well in Galactic dwarf galaxies \cite{Fattahi}. Also, baryons usually dominate the inner regions of galaxies. The error of using the NFW profile in this model is very small.

Furthermore, as mentioned above, some studies indicate that the observed dark matter-baryon relations can be explained by the Modified Newtonian Dynamics (MOND) \cite{McGaugh,Milgrom}. It is interesting to note that many studies connecting dark matter and baryons involve a characteristic universal constant $a_0 \sim 10^{-10}$ m/s$^2$. For example, Gentile et al. (2009) \cite{Gentile} discover that the mean dark matter surface density within one dark halo scale-length is constant. The constant is proportional to a universal gravitational acceleration $g_{dark}=3.2^{+1.8}_{-1.2} \times 10^{-11}$ m/s$^2$. For the radial acceleration relation, the empirical expression involves a constant $g_0=1.20 \times 10^{-10}$ m/s$^2$ \cite{McGaugh}. In our model, we suggest that this value corresponds to the term $g_{d0}=4 \pi G\rho_sr_s \approx (2-4) \times 10^{-10}$ m/s$^2$ that exists in the analytic expressions. We show that this term depends on the total dark matter mass slowly ($\rho_sr_s \propto M_{dt}^{0.2}$). Therefore, the range of this term is very small so that it seems to be a universal constant for all galaxies. 

To conclude, the observed dark matter-baryon relations can be explained by the CDM model. The analytic expressions derived can give excellent agreements with the observations and explain why the resulting relations are so tight. 

\section{acknowledgements}
This work is supported by a grant from The Education University of Hong Kong (Project No.:RG4/2016-2017R).

\end{document}